\documentclass[11pt]{article}
\usepackage[a4paper]{geometry}
\usepackage{amssymb}
\usepackage[normalsize,normal,sc]{caption}

\usepackage[latin1]{inputenc}

\def\FLUCTUN{.024234}
\def\FLUCTDEUX{9.0054\,10^{-5}}
\def\PUNSTAR{.14846}
\def\PDEUXSTAR{.07929}

\def \eref#1{(\ref{#1})}
\def \j{j^\star}
\def \no{\noindent}
\def \dis{\displaystyle}
\def \l{\left}
\def \r{\right}
\def \B{\Big}

\def \beq{\begin{equation}}
\def \eeq{\end{equation}}
\def \be{\begin{eqnarray*}}
\def \ee{\end{eqnarray*}}
\def \ben{\begin{eqnarray}}
\def \een{\end{eqnarray}}
\def \mm{\medskip}
\def \bb{\bigskip}
\def \s{\smallskip}
\def \qE{\mathbb{E}}
\def \qZ{\mathbb{Z}}
\def \qP{\mathbb{P}}
\def \BO{\mathcal{O}}
\def \equaldef{ \stackrel{\tiny{\mbox{def}}\normalsize}{=} }
\newcommand{\bm}[1]{\mbox{\boldmath $ #1 $}}

\def\bibfmta#1#2#3#4{{#1}, {#2}, \textit{#3}, #4.}
\def\bibfmtb#1#2#3#4{{#1}, \textit{#2}, {#3}, #4.}

\begin{document}

\newtheorem{fig}{\hspace{2cm} Figure}
\newtheorem{lem}{Lemma}
\newtheorem{defi}{Definition}
\newtheorem{definot}{Definition and Notation}
\newtheorem{pro}{Proposition}
\newtheorem{rem}{Remark}
\newtheorem{theo}{Theorem}
\newtheorem{cor}{Corollary}
\newenvironment{conj}{\newline\bf Conjecture:\sl}{\rm}
\newenvironment{proof}{\no \textbf{Proof:} }{\hfill $\square$}

\title{\textbf{Quasi-Optimal Energy-Efficient} \\[.1\baselineskip]
\textbf{Leader Election Algorithms in Radio Networks}
}

\author{Christian \textsc{Lavault}\thanks{Corresponding author:
LIPN, CNRS UMR 7030, Universit\'e Paris~13 -- 99, av. J.-B. Cl\'ement %
93430 Villetaneuse, France.\ \ E-mail: lavault@lipn.univ-paris13.fr}
\and
Jean-Fran\c{c}ois \textsc{Marckert}\thanks{CNRS, LaBRI Université
    Bordeaux 1, 351, cours de la Lib\'eration 33405 Talence cedex,
    France,\ E-mail: marckert@labri.fr}
\and
Vlady \textsc{Ravelomanana}\thanks{LIPN, UMR CNRS 7030,\ %
Universit\'e Paris 13, France.\ E-mail:\ vlad@lipn.univ-paris13.fr}
}
\date{\empty}
\maketitle

\begin{abstract}
Radio networks (RN) are distributed systems (\textit{ad hoc networks})
consisting in $n \ge 2$ radio stations. Assuming the number $n$ unknown,
two distinct models of RN without collision detection (\textit{no-CD})
are addressed: the model with \textit{weak no-CD} RN and the one with
\textit{strong no-CD} RN.
We design and analyze two distributed leader election protocols,
each one running in each of the above two (no-CD RN) models, respectively.
Both randomized protocols are shown to elect a leader within $\BO(\log{(n)})$
expected time, with no station being awake for more than $\BO(\log{\log{(n)}})$
time slots (such algorithms are said to be \textit{energy-efficient}).
Therefore, a new class of efficient algorithms is set up that match
the $\Omega(\log{(n)})$ time lower-bound established by Kushilevitz
and Mansour in~\cite{Mansour}.
\end{abstract}

\section{Introduction}
Electing a leader is a fundamental problem in distributed systems
and it is studied in a variety of contexts including radio
networks~\cite{CHLEBUS}. A radio network (RN, for short) can be
viewed as a distributed system of $n$ radio stations with no
central controller. The stations are bulk-produced, hand-held
devices and are also assumed to be indistinguishable: no
identification numbers (or IDs) are available. A large body of
research has already focused on finding efficient solutions to
elect one station among an $n$-stations RN under various assumptions
(see e.g.~\cite{CHLEBUS,Mansour,Willard}). It is also assumed that
the stations run on batteries. Therefore, saving battery power
is important, since recharging batteries may not be possible
in standard working conditions. We are interested in designing
\textit{power-saving} protocols (also called \textit{energy-efficient}
protocols). The present work is motivated by various applications
in emerging technologies: from wireless communications, cellular
telephony, cellular data, etc., to simple hand-held multimedia services.

\bb \no \textbf{The models.}\ As customary, the time is assumed to be
slotted, the stations work synchronously and they have no IDs available.
No \textit{a priori} knowledge is assumed on the number $n\ge 2$
of stations involved in the RN: neither a (non-trivial) lower-bound
nor an upper-bound on $n$.
During a time slot, a station may be awake or sleeping. When sleeping,
a station is unable to listen or to send a message. Awake stations communicate
globally, i.e.~the underlying graph is a clique, by using a unique
radio frequency channel with \textit{no collision detection}
(\textit{no-CD} for short) mechanism. In each time slot, the status
of the unique channel can be in one of the following two states: \\
$\bullet$ either \textsc{SINGLE}: there is exactly one transmitting station, \\
$\bullet$ or \textsc{NULL}: there is either no station or more
than two ($\ge 2$) broadcasting stations.

When the status is NULL, each listening station hears some noise and can
not decide whether 0 or more than 2 station are broadcasting.
When the status is SINGLE, each listening station hears clearly the message
sent by the unique broadcasting station.

\mm In the \textit{weak} no-CD model of RN, during a time slot each awake station
may \textit{either} send (broadcast) a message \textit{or} listen to the channel,
\textit{exclusively}.
By contrast, in  the \textit{strong} no-CD model of RN, both operations
can be performed \textit{simultaneously} by each awake station, during a time slot.
Hence, in the strong no-CD model, when exactly one station sends at time slot $t$,
then all the stations that listen at time $t$, transmitter included, eventually
receive the message. In the literature, the no-CD RN usually means the strong model
of RN, see e.g.~\cite{Mansour,OLARIU}. In the weak no-CD case, such a transmitting
station is not aware of the channel status.

Such models feature concrete situations; in particular, the lack
of feedback mechanism experiences real-life applications (see
e.g.~\cite{NOCOLLISION}). Usually, the natural noise existing within
radio channels makes it impossible to carry out a message collision
detection.

\bb \no \textbf{Related works.}\ The model of RN considered is the broadcast
network model (see e.g.~\cite{CHLEBUS}). In this setting, the results
of Willard~\cite{Willard}, Greenberg \textit{et al.}~\cite{PHILIPPE-ACM}
(with collision detection) or Kushilevitz and Mansour~\cite{Mansour}
(no-CD) for example, are among the most popular leader election protocols.

In such a model, Massey and Mathys~\cite{NOCOLLISION} serves as a global
reference for basic conflict-resolution based protocols. Previous researches
on RN with multiple-access channel mainly concern stations that are kept
awake during all the running time of a protocol even when such stations
are the ``very first losers'' of a coin flipping game algorithm~\cite{Helmut}.
In~\cite{Polonais}, the authors design an energy-efficient protocol
(with $o(\log\log(n))$ energy cost) that approximate $n$ up to a constant
factor. However, the running time achieved is $\BO\l(\log^{2+\epsilon}(n)\r)$
in strong no-CD RN. Also, an important issue in RN is to perform distribution
analysis of various randomized election protocols as derived
in~\cite{Szpankowski-Fill-Mahmoud,Szpankowski-Janson} for example.

\bb \no \textbf{Our results.}\ Two leader election protocols are
provided in the present paper. The first one (Algorithm~1) is
designed for the \textit{strong} no-CD model of RN, while the second
one (Algorithm~2) is designed for the \textit{weak} no-CD model of
RN. Both are double-loop randomized algorithms, which use a simple
coin-tossing procedure (\textit{rejection algorithms}). Our leader
election protocols achieve a (quasi) optimal $\BO(\log{n})$ average
running time complexity, with no station in the RN being awake for
more than $\BO(\log{\log{(n)}})$ time slots.
Indeed, both algorithms match the $\Omega(\log{n})$ time lower-bound
established in~\cite{Mansour} and also allow the stations to keep
sleeping most of the time. In other words, each algorithm greatly reduces
the total awake time slots of the $n$ stations: shrinking from the usual
$\BO(n\log{n})$ down to $\BO(n\log\log{n})$, while their expected time
complexity still remains $\BO(\log{n})$. Our protocols are thus
``energy-efficient'' and suitable for hand-held devices working with batteries.

Besides the algorithms use a parameter $\alpha$ which works as a flexible
regulator. By tuning the value of $\alpha$
the running time ratio of each protocol to its energy consumption
may be  adjusted~: the running time and the
awake duration are functions of $\alpha$.

Furthermore, the design of Algorithms~1 and 2 suggests that within
both the weak and the strong no-CD RN, the average time complexity
of each algorithm only differs from a slight constant factor.
Also, our results improve on~\cite{OLARIUCONF}.

\bb \no \textbf{Outline of the paper.}\ In Section~2, we present
Algorithm~1 and Algorithm~2, and the main complexity results are given:
Theorem~\ref{Theorem-ALGO1} and Theorem~\ref{Theorem-ALGO2},
corresponding to Algorithm~1 and Algorithm~2.
Section~3 and 4 are devoted to the analysis of both algorithms, by means
of tight asymptotic techniques.

\section{Algorithms and Results}
Both algorithms rely on the intuitive evidence that all stations
must be awake altogether within a sequence of predetermined time
slots in order to be informed that the election succeeds of fails.

\mm To realize the election a first naive idea is to have
stations using probabilities $1/2$, $1/4$,\ldots to wake up and broadcast.
This solution is not correct since with probability $>0$ never a station
ever broadcasts alone. In order to correct the failure we will plan
an unbounded sequence of rounds with predetermined length.
Awake time slots are programmed at the end of each rounds in order
to allow all stations to detect the possible termination of the session.

\bb \no In the following, we let $\alpha$ denote a real (tuning)
parameter value, which is required to be $> 1$ (this is explained
in Remarks~\ref{alpha-cq1} and~\ref{alpha-cq2}).

\subsection{Algorithm 1} \label{algo1}
In Algorithm 1 the stations work independently. Given a round $j$ in the
outer loop (repeat-until loop), during the execution time of the inner loop
each station randomly chooses to sleep or to be awake~:  in this last case
it listens and broadcast, simultaneously. If a unique station is broadcasting,
this station knows the status of the radio channel; if the status is SINGLE,
it becomes a \textit{candidate}. At the end of round $j$, every station
wakes up and listens to the channel and the candidates broadcast.
If there is a single candidate, the status is SINGLE again, and this candidate
is elected. Every listening station knows the channel status, and is informed
that the election is done. Otherwise, the next round begins.

\vskip 0.5cm
\no \hrule \vspace{0.05cm}
\indent $round \leftarrow 1$; \\
\indent \textbf{Repeat}

\mm
\indent \indent \textbf{For}\ $k$ from 1 to $\lceil \alpha^{round} \rceil$\ %
            \textbf{do} \hfill /* probabilistic phase */ \\
\indent \indent \indent$\left[\begin{array}{l}\textrm{Each station wakes up %
                                    independently with probability}\ 1/2^{k}. \\
\textrm{An awake station listens and broadcasts} \\
\textrm{\textbf{If}\ a unique station broadcasts\ \textbf{Then}\ %
            it becomes a \textit{candidate} station}\ \textbf{EndIf}\end{array}\right.$ \\
\smallskip
\indent \indent \textbf{EndFor} \hfill /* deterministic phase */

\medskip
\indent \indent $\left[\begin{array}{l}\textrm{At the end of each round, %
                                        all stations wake up, listen,} \\
\textrm{ and in addition all \textit{candidate} stations broadcast;}\\
\textrm{ \textbf{If}\ there is a unique \textit{candidate}\ \textbf{then}\
                it is \textit{elected}\ \textbf{EndIf}};
         \end{array}\right.$\medskip\\
\indent  $round \leftarrow round + 1$;

\medskip
\indent \textbf{until a station is elected}
\vspace{0.25cm} \hrule
\begin{center}
\textbf{Algorithm 1. Leader election protocol for strong no-CD RN}
\end{center}

\bb The brackets in both algorithm represent the actions that take place
in one time slot. Notice that the content of the broadcasting message
is not specified, since it has no importance. The status ``candidate''
is valid for one round duration only.

\begin{definot}
For the sake of simplicity, the following notations
are used throughout the paper. We let
\beq \label{JN_STAR}
j^{\star} \equiv j^{\star}(n) \;\equaldef \;\lceil \log_{\alpha}\log_{2}(n) \rceil .
\eeq
Let also $\mathcal{C} : \mathbb{R}^2\to \mathbb{R}$ defined by

\begin{equation} \label{CC}
\mathcal{C}(x,y) \equaldef \;\frac{xy^{3}}{(y-1)\B(1-y (1-x)\B)}.
\end{equation}
\end{definot}
We have

\begin{theo} \label{Theorem-ALGO1}
Let $\alpha\in(1,1.1743\ldots)$ and $p_1^{\star} = \PUNSTAR\ldots$
On the average, Algorithm~$1$\ elects a leader in at most\
$\mathcal{C}({p_1^{\star}},\alpha) \log_2{(n)} \,+\, \BO(\log\log n)$\
time slots, with no station being awake for more than
$2 \log_{\alpha}\log_{2}(n)\,\B(1 + o(1)\B)$\ mean time slots.
\end{theo}

\subsection{Algorithm 2} \label{algo2}
In the case of the weak no-CD model of RN a potential candidate
alone cannot be aware of its status since it cannot broadcast and
listen \textit{simultaneously}. The awake stations choose to
 broadcasts or (exclusively) to listen.
In the inner loop of Algorithm 2 any exclusively broadcasting station
is called an \textit{initiator}; when there is a unique initiator,
a station that is listening to the channel, hears clearly the message.
Such a station is said to be a \textit{witness}. Witnesses are intended
to acknowledge an initiator in the case when there is exactly one initiator.

\vskip 0.5cm
\no \hrule \vspace{0.05cm}
\indent $round \leftarrow 1$; \\
\indent \textbf{Repeat}

\medskip
\indent \indent \textbf{For}\ $k$ from 1 to $\lceil \alpha^{round} \rceil$\ %
\textbf{do} \hfill /* probabilistic phase */

\smallskip
\indent \indent \indent $\left[\begin{array}{l}\textrm{Each station wakes up %
                                independently with probability}\ 1/2^k; \\
\textrm{With probability}\ 1/2, \textrm{each awake station \textit{either} broadcasts} \\
\textrm{the message}\langle k\rangle\ \textrm{\textit{or} listens, exclusively;} \\
\textrm{If there is a unique initiator, the status is SINGLE;} \\
\textrm{The witness(es) record(s) the value}\ \langle k\rangle\
                                \textbf{EndIf}\end{array}\right.$

\smallskip
\indent \indent \textbf{EndFor} \hfill /* deterministic phase */

\bigskip
\indent \indent $\left[\begin{array}{l}\textrm{At the end of each round, %
                                                all stations wake up;} \\
\textrm{All witnesses forward the recorded message;} \\
\textrm{\textbf{If}\ the status is SINGLE} \\
\textrm{there is a unique witness, the other stations receive the message}\ \langle k\rangle \\
\textrm{\textbf{Then}\ the unique initiator of the message}\ %
                    \langle k\rangle\ \textrm{is \textit{elected}}\end{array}\right.$ \\~\\
\indent \indent \indent $\left[\begin{array}{l}\textrm{and replies %
        to all the stations to advise of its status}\ \textbf{EndIf}\end{array}\right.$ \\

\smallskip
\indent \indent $round \leftarrow round + 1$;

\medskip
\indent \textbf{until a station is elected.}
\vspace{0.25cm}
\hrule
\begin{center}
\textbf{Algorithm 2. Leader election protocol for weak no-CD RN}
\end{center}

\bb
During the last deterministic phase all stations wake up. An election
takes place in a round if this round is the first when there is a unique
time slot in which there remain exactly a unique initiator and a unique witness.
When a station is a witness, it continues to behave as it were not.
Hence, a witness can be a witness twice or more, or even be an initiator.
This is obviously not optimal but simpler for the analysis.

We have
\begin{theo} \label{Theorem-ALGO2}
Let $\alpha\in(1,1.0861\ldots)$ and $p_2^{\star} = \PDEUXSTAR\ldots$
On the average, Algorithm~$2$\ elects a leader in at most\
$\mathcal{C}(p_2^{\star},\alpha) \log_2{(n)} \,+\, \BO(\log\log n)$\
time slots, with no station being awake for more than
$5/2 \,\log_{\alpha}\log_{2}(n)\,\B(1 + o(1)\B)$\ mean time slots.
\end{theo}

\section{Analysis of Algorithm 1}
Assume that Algorithm 1 begins with its variable $round = j$,
and let $p_j$ be the probability that one station is \textit{elected}
at the end of this round $j$. One may also view $p_j$ has the conditional
probability that an election occurs at round $j$ knowing that it did not
occur before. Within that round, that is for $k$ ranging from $1$ to
$\lceil\alpha^{j} \rceil$, every station decides to broadcast
with the sequence of probabilities $(1/{2^k})_{1\le k\le \lceil \alpha^{j}\rceil}$.

\mm The probability $\rho_{(i,n)}$ that there is exactly one candidate
for $k = i$ in a round is $\dis \rho_{(i,n)} =\; \frac{n}{2^i}\l(1-\frac{1}{2^i}\r)^{n-1}$.
We then have
\ben
p_{j} & = & \sum_{k=1}^{\lceil \alpha^{j} \rceil} \rho_{(k,n)} %
\prod_{{i=1 \atop i\neq k}}^{\lceil \alpha^{j}\rceil} \l(1 - \rho_{(i,n)}\r)%
\;=\;  \sum_{k=1}^{\lceil \alpha^{j} \rceil} \rho_{(k,n)} \;\l(1-\rho_{(k,n)}\r)^{-1} %
\,\prod_{i=1}^{\lceil \alpha^{j} \rceil} \l(1 - \rho_{(i,n)}\r)
\een
and $p_j = t_j\, s_j$, with
\beq \label{GERARDO}
t_j\; \equiv \; \sum_{m=0}^{\infty} \sum_{k=1}^{\lceil \alpha^{j}\rceil}\l(\rho_{(k,n)}\r)^{(m+1)}\ %
\quad \mbox{and}\ \quad s_j\; \equiv \;\prod_{i=1}^{\lceil \alpha^{j}\rceil}(1 - \rho_{(i,n)}).
\eeq

\mm \no Note that $s_j$ represents the probability of having no candidate in the $j$th round.

\begin{rem} \label{CRUCIAL}
Simple considerations show that when $2^{\alpha^{j}} \ll n$,
the probability $s_{j}$ is close to $1$ (and is ``far from'' $1$
when $2^{\alpha^{j}}\gg n$).
This remark explains the occurrences of the crucial values $n/{2^{\alpha^{j}}}$
and the definition of $j^{\star}$ in Eq.~\eref{JN_STAR}.
\end{rem}

\subsection{Two Lemmas for a Lower Bound on $\bm{p_{j}}$}
Lemma~\ref{Lemma-LOWBOUND-SJ} provides a lower bound
on the finite product denoted by $s_j$ in Eq.~(\ref{GERARDO}).

\begin{lem} \label{Lemma-LOWBOUND-SJ}
Let $j\equiv j(n)$ such that $j(n)\to +\infty$ and $n/2^{\alpha^{j}}\to 0$.
We have
\[\liminf_{n} s_j\; \ge \,.1809\ldots.\]
\end{lem}

\begin{proof}
We divide the product $s_{j}$\ in (\ref{GERARDO}) as follows:
for any $r\in\{2,\dots, \lceil \alpha^j\rceil\}$,
\begin{equation}\label{eq23}
s_{j} =\; \prod_{i=1}^{r-1}
(1 - \rho_{(i,n)}) \times \prod_{i=r}^{\lceil \alpha^j\rceil} (1 - \rho_{(i,n)}).
\end{equation}

Since for any $a\in[0,1]$ and $n\ge 1$, $(1-a)^n \le e^{-an}$,
we have for any $i\ge 1$,
\ben \label{ineq1}
1 - \rho_{(i,n)} \;\ge \; %
1 - \frac{n}{2^i} \exp{\l(-\frac{n-1}{2^i} \r)} \;\ge \; %
1 - \frac{n}{2^i} \exp{\l(-\frac{n}{2^i} \r)}\ e^{1/2};
\een
and more precisely, for $i\in\{r,\dots, \lceil \alpha^j\rceil\}$,
with $r$ large enough,
\ben\label{ineq2}
1 - \rho_{(i,n)} \;\ge \; %
1 - \frac{n}{2^i} \exp{\l(- \frac{n}{2^i} \r)} \exp{\l(\frac{1}{2^r} \r)} \;\ge \; %
1 - \frac{n}{2^i} \exp{\l(- \frac{n}{2^i} \r)} \l(1 + \frac{2}{2^r} \r).
\een
Now we let $\dis r \equiv r(n) = \lfloor 1/2\log_2{(n)}\rfloor$,
ensuring that $2^r/\!\sqrt{n}$\ is bounded. A simple bounding argument
in Eq.~\eref{ineq1} shows that $\dis \prod_{i=1}^{r-1} (1 - \rho_{(i,n)})\to 1$\
when $n\to \infty$. \\
Thus, using also inequation~\eref{ineq2}, Eq.~\eref{eq23} yields
\ben
\liminf_n s_{j} & \ge & \exp{\l( \sum_{i=r}^{\lceil \alpha^j\rceil} %
\ln{\l(1-\l(1+\frac{2}{2^{r}}\r) \,\frac{n}{2^i}\, \exp{\l(-\frac{n}{2^i}\r)}\r)}\r)} \cr
& & \cr
& \ge & \exp{ \l(-\sum_{m\ge 1}\, \frac{\l(1 + \frac{2}{2^{r}}\r)^m}{m}\ %
\sum_{i=1}^{\lfloor\alpha^j\rfloor} \frac{n^m}{2^{i m}}\, \exp{\l(-\frac{nm}{2^i}\r)}\r)}.\ \ \ \
\label{sumsinexp}
\een
Denote by $A_n$ the right hand side of the inequality. Computing the second sum
(inside the exponential in $A_n$) is completed through asymptotic
approximations using the Mellin transform in Lemma~\ref{LEMME-MELLIN1}
given in Appendix~\ref{TECHLEM} (this is possible since $n/2^{\alpha^j}\to 0$
by assumption). By Lemma~\ref{LEMME-MELLIN1},
\ben
A_n & \ge & \exp{\l(o(1) -\, \sum_{m\ge 1} \frac{(1 + \frac{2}{2^{r}})^m m!}{m^{m+2}\ln{2}} %
\;-\; \frac{(1+\frac{2}{2^{r}})^m}{m^{2}\,2^m}\, \sup_{\nu}|U_{\nu}(\log_2{n})|\r)},
\label{SummationonMinexp}
\een
where $\sup_{\nu}|U_{\nu}(z)| < \FLUCTUN$ for all $z$, as stated
in Lemma~\ref{LEMME-MELLIN1} (Appendix~\ref{TECHLEM}).

\no Therefore, the third term in Eq.~(\ref{SummationonMinexp})
is bounded from below by
\be
\exp{\Big(-\FLUCTUN \,\sum_{m\ge 1} \frac{1}{m^{2}}\Big)} \;= .96092\ldots
\ee
Next, the second term within the exponential in Eq.~(\ref{SummationonMinexp})
expresses as
\ben
\sum_{m=1}^{\infty} \frac{ \l(1 + \frac{2}{2^r} \r)^m m!}{(\ln 2)\,m^{m+2} } %
& \le & \sum_{m=1}^{n^{1/6}} \frac{\l(1 + \frac{2}{2^r} \r)^m m!}{(\ln 2)\,m^{m+2}}\ %
+ \sum_{m=n^{1/6}+1}^{\infty} \frac{2^m m!}{(\ln 2)\,m^{m+2}}\,. \label{splittedSum}
\een
When $n$ (and so $r$) tends to $+\infty$, the first sum on
the right hand side of Eq.~(\ref{splittedSum}) converges to
$\dis \sum_{m\ge 1} \frac{m!}{(\ln 2)\,m^{m+2}} = 1.6702\ldots$,
as a consequence of Lebesgue's dominated convergence theorem.
And the second term goes to 0, as derived from Stirling formula. \\
Finally, combining all these facts we obtain the announced lower bound on $s_j$.
\end{proof}

\mm
The following Lemma~\ref{0.148} provides a lower bound on $\dis \liminf_n p_{j}$.

\begin{lem} \label{0.148}
Let $j\equiv j(n)$ such that $j(n)\to +\infty$ and $n/2^{\alpha^{j}}\to 0$. We have
$$\liminf_{n} p_{j} > p_1^{\star} = \PUNSTAR\ldots$$
\end{lem}

\begin{proof}
Since $(1-x) \ge \exp{(-2x)}$\ for $x\in [0,1/2]$, we get the following

\mm \no $\dis \sum_{k=1}^{\lceil \alpha^{j}\rceil} (\rho_{(k,n)})^{(m+1)}$
\vspace{-.3cm}
\be
\ge & & \!\!\sum_{k=1}^{\lceil \alpha^{j}\rceil} \l(\frac{n}{2^k}\, %
\exp{\l(-\frac{n-1}{2^{k-1}}\r)}\r)^{(m+1)}
\;\ge \;\;\sum_{k=1}^{\lceil \alpha^{j}\rceil} \frac{1}{2^{m+1}}\, %
\l(\frac{n}{2^{k-1}}\r)^{m+1}\! \exp{\l(-\frac{n(m+1)}{2^{k-1}}\r)} \cr
& & \cr
= & & \!\!\sum_{k=1}^{\lceil \alpha^{j}\rceil} \frac{1}{2^{m+1}}\, %
\l(\frac{n}{2^{k}}\r)^{m+1} \exp{\l(-\frac{n(m+1)}{2^{k}}\r)} \cr
& & \qquad +\;\; \frac{1}{2^{m+1}}\,\l(n^{m+1} \exp{\B(-n(m+1)\B)} \;-\; %
\l(\frac{n}{2^{\lceil \alpha^{j}\rceil}}\r)^{m+1}\!\! %
\exp{\l(-\frac{n(m+1)}{2^{\lceil \alpha^{j}\rceil}}\r)}\r) \cr
& & \cr
= & & \!\!\frac{1}{2^{m+1}}\; \frac{(m+1)!}{(m+1)^{m+2}\,\ln{2}} \;+\; %
\frac{U_m \B(\log_{2}(n)\B)}{(m+1)^{2}\,4^{m+1}} \;+\; %
\BO\l(\frac{n}{2^{\alpha^j+m+1}} + \frac{1}{n\,2^{m+1}}\r). \label{NELSON}
\ee
By assumption, $n/2^{\alpha^{j}} \to 0$. So, the latter expression comes
from Lemma~\ref{LEMME-MELLIN1}, where the term $U_m(z)$ is defined in
Eq.~(\ref{EQ:UMZPHI}). Now, summing on $m$ in Eq.~(\ref{GERARDO}) we derive
\be
t_{j} \ge \BO\l( \frac{n}{2^{\alpha^j}}+\frac{1}{n} \r) \;+\; %
\sum_{m=1}^{\infty} \frac{m!}{2^m\, m^{m+1}\ln{2}} \;-\; \sum_{m=1}^{\infty} %
\frac{\sup_{\nu}|U_{\nu}(\log_2{n})|}{m^2\;4^{m}}\,.
\ee
By Eq.~\eref{fluc}, numerical evaluation of the sums (with Maple)
give $\dis \liminf_n t_{j} \ge .82092\ldots$

\no Finally, by using both lower bounds on $s_{j}$ and $t_{j}$
(in Lemma~\ref{Lemma-LOWBOUND-SJ} and in the above proof, resp.)
we obtain the desired lower bound on $\dis \liminf_{n} p_{j}$:
$p_1^{\star}$ = \PUNSTAR\ldots
\end{proof}

\subsection{Proof of Theorem \ref{Theorem-ALGO1}}
To prove Theorem \ref{Theorem-ALGO1}, we use the following
Lemma~\ref{GEOMETRIC_INEQUALITIES} that allows us to control
the mean of the quantities of interests, whenever having a bound
on the probability of success in each round.
\begin{lem} \label{GEOMETRIC_INEQUALITIES}
Let $(X_i)_{i\ge 1}$ and $(Y_i)_{i\ge 1}$ be two sequences of
independent Bernoulli random variables, denoted by $B(P_i)$ and
$B(Q_i)$, respectively, and such that $P_i\ge Q_i$ for any $i$.
By definition,
$$\qP(X_i=1) = 1 - \qP(X_i=0) = P_i\ \quad \textrm{and}\ %
\quad \qP(Y_i=1) = 1 - \qP(Y_i=0) = Q_i.$$

\no Let $H = \inf\{j, X_j = 1\}$ and $K = \inf\{j, Y_j = 1\}$,
which may be regarded as a first success in each sequence $X_i$
and $Y_i$ (resp.). Then, the ``stochastic inequality'' $H\le_S K$
holds: for any non-negative integer $k$, $\qP(H \le k)\ge \qP(K\le k).$

\no Moreover, for any non-decreasing function $f$,
\begin{equation} \label{eq:stochasticexpectation}
\qE(f(H))\le \qE(f(K)).
\end{equation}
\end{lem}

\begin{proof}
The first part of the Lemma can be proved by constructing a probability
space $\Omega$ in which the sequences of r.v. $(X_i)$ and $(Y_i)$ ``live''
and in which, for each $\omega\in\Omega$, $X_i(\omega)\geq Y_i(\omega)$.
Since for any $\omega$, $K(\omega)\geq H(\omega)$, the stochastic order
is a simple consequence of this almost sure order on $\Omega$.
Next, for any nondecreasing function $f$, $f(K(\omega))\ge f(H(\omega))$
also holds almost surely and whence Eq.~(\ref{eq:stochasticexpectation}).
\end{proof}

\mm We come back to the proof of Theorem~\ref{Theorem-ALGO1}.
It is straightforward to check that, when $n \to \infty$,
$n/2^{\alpha^{(\j+1)}}\to 0$. According to Lemma~\ref{0.148},
if $n$ is large enough,
\ben \label{MARCEL}
p_j\;\ge \;p_1^{\star}\,\mathbb{I}_{j\ge \j+1}
\een

As a consequence, let $N_1$ denote the number of rounds in Algorithm~1
and $N_1' = j^{\star} + G$ where $G$ is a geometric random variable
with parameter $p_1^{\star}$, then
\ben \label{ST}
N_1\,\le_S \,N'_1,
\een
($N_1$ is smaller with respect to the stochastic order than $N_1'$).

\no We recall that the geometrical distribution with parameter $p$
is given by $p(1-p)^k$ for any $k\ge 1$; this is the law of the first success
in a sequence of independent Bernoulli random variables with parameter $p$.

\mm To show Eq.~(\ref{ST}), notice first that $N_1$ has the same
distribution as $\inf \{i, X_i=1\}$, where the $(X_i)$ are independent
and $X_i$ is $B(p_i)$-distributed. \\
Next, $N'_1 = \inf \{i, Y_i=1\}$, where $Y_i$ is $B(q_i)$-distributed for
$q_i = p_1^{\star}\,\mathbb{I}_{j\ge \j+1}$. Indeed, the first $j^{\star}$
trials fail and afterwards, each trial results in a success with probability
$p_1^{\star}$. \\
Finally, Eq.~(\ref{MARCEL}) and Lemma~\ref{eq:stochasticexpectation} allow
to conclude,
$$\qE(N_1) \;\le \; \qE(N_1') \;=\; j^{\star} \;+\; 1/p_1^{\star} %
\;=\; \log_{\alpha}\log_{2}(n) \;+\; \BO(1).$$

Let $T_1 \equiv T_1(n)$ be the time needed to elect a leader in
Algorithm 1. Since $N_1'$ is larger than $N_1$ for the stochastic
order and since $r\mapsto \sum_{i=1}^{r} \lceil \alpha^i \rceil $ is
non-decreasing, by Lemma~\ref{GEOMETRIC_INEQUALITIES},
\ben
\qE(T_1) & = & \qE\l(\sum_{j=1}^{N_1}\lceil \alpha^j \rceil\r) %
\;\le\;\qE\l(\sum_{j=1}^{N_1'}\lceil \alpha^j \rceil\r) \;\le\; %
\sum_{k=1}^{+\infty}\, \sum_{j=1}^{j^{\star}+k}(1+\alpha^j) %
p_1^{\star}(1 - p_1^{\star})^{k-1} \cr
& & \cr
& \le & {\cal C}(p_1^{\star},\alpha) \log_2 (n) \;+\; \BO(\log\log n). \label{doublesum}
\een

\mm \no During a round, the mean number of awake time slots for a given
station is at most 2. (Two at the end, one from the contribution of
$\sum_k 1/2^k$, $1/2^{k}$ being the probability to be awake at time slot $k$.)

Since the number of rounds is smaller than, say: $\log_{\alpha}\log_{2}(n) + \log\log\log(n)$
with probability going to 1, we get the announced result.  \hfill $\square$

\begin{rem} \label{alpha-cq1}
It is easily seen that the algorithm and the convergence of the double
sum in Eq.~(\ref{doublesum}) (resp.) require the conditions $\alpha > 1$
and $\alpha (1 - p_1^{\star}) < 1$, with $p_1^{\star} = \PUNSTAR$ (resp.).
The value of $\alpha$ may thus be chosen in the range $(1,1.743\ldots)$,
so as to achieve a tradeoff between the average execution time of the
algorithm and the global awake time.
Thus, the minimum value of the constant ${\cal C}(p_1^{\star},\alpha)$
is ${\cal C}(p_1^{\star},\widetilde{\alpha})\simeq  29.058\ldots$,
with $\widetilde{\alpha} = 1.0767..\ldots$
\end{rem}

\section{Analysis of Algorithm 2}
Two awake stations are needed in Algorithm~2: the one is only sending
(the initiator) and the other is listening. The probability $p'_{j}$ that
one station is elected in round $j$ expresses along the same lines as in
Eq.~(\ref{GERARDO}) with the corresponding probability. The probability
of having exactly one initiator and one witness for $k = j$ in a round is
$$q_j^n\equiv \frac{1}{2}\frac{{n\choose 2}}{4^j}\l(1-\frac{1}{2^j}\r)^{n-2}.$$
Hence $p'_{j} = t'_{j}\, s'_{j}$, where
\ben \label{PAOLO}
t'_{j} & \equiv &  \sum_{k=1}^{\lceil \alpha^{j} \rceil} q_{(k,n)}\l(1-q_{(k,n)}\r)^{-1} %
\quad \mbox{and}\ \quad  s'_j\equiv  \prod_{i=1}^{\lceil \alpha^{j} \rceil}(1-q_{(i,n)}).
\een
All computations are quite similar to those in the proof of Theorem~\ref{Theorem-ALGO1}.
Again, the value of $p'_j$ in Eq.~(\ref{PAOLO}) is obtained by asymptotic
approximation (see Lemma~\ref{LEMME-MELLIN2} in Appendix~\ref{TECHLEM}).

\s \no First,
\be
s'_{j} & \ge & \prod_{i=1}^{\lceil \alpha^{j}\rceil} \l(1 \,-\; \frac{1}{2}\, %
\frac{{n\choose 2}}{4^i}\, \exp{\l(-\frac{n-2}{2^i}\r)}\r),\ \qquad \mbox{since}\ %
\forall x \in \l[0,1\r]\ \ 1 - x \le e^{-x}\,, \cr
& & \cr
& \ge & \prod_{i=1}^{\lceil \alpha^{j}\rceil} \l(1 \,-\; \frac{n^2}{4^{i+1}}\, %
\exp{\l(-\frac{n-2}{2^i} \r)}\r),\qquad \quad \mbox{since}\ {n\choose 2}\le \frac{n^2}{2}\,, \cr
&  & \cr
& \ge & \prod_{i=1}^{\lceil \alpha^{j} \rceil} \l(1 \,-\, \frac{n^2}{4^{i}}\, %
\exp{\l(-\frac{n}{2^i}\r) \frac{e}{4}}\r) \;\ge \; %
\prod_{i=1}^{\lceil \alpha^{j}\rceil} \l(1 \,-\; \frac{n^2}{4^{i}}\, %
\exp{\l(-\frac{n}{2^i}\r)}\r).
\ee

\no Using this latter lower bound in Lemma~\ref{LEMME-MELLIN2} yields
\be
s'_{j} & \ge & \exp{\l(-\sum_{m=1}^{\infty} \sum_{i=1}^{\lceil \alpha^{j} \rceil} %
\frac{1}{m}\, \frac{n^{2m}}{4^{im}}\, \exp{\l(-\frac{nm}{2^i}\r)}\r)} \cr
& & \cr
& \ge & \exp{\l(o(1) -\; \sum_{m=1}^{\infty} \frac{1}{m}\, \frac{(2m-1)!}{m^{2m+1}\;\ln2} %
\;-\; \sum_{m=1}^{\infty} \l(\frac{1}{m^3}\, \sup_{\nu}|V_{\nu}(\log_2{n})|\r)\r)},
\ee
with the following value of the lower bound on $s'_j$ (computed with Maple):
\[\liminf_n s'_j\ge .19895\ldots\]

\mm \no Next, $1 - x^2\ge e^{-x}$ when $x$ is close to $0$ and so
$\dis {n\choose 2}\ge \frac{n^2}{2} \exp{\l(-\frac{1}{\sqrt{n}}\r)}$;
whence $t'_{j}$ also is bounded from below as follows,
\be
t'_{j} & \ge & \sum_{k=1}^{\lceil \alpha^{j}\rceil} \sum_{m=1}^{\infty} \l(\frac{1}{2}\, %
\frac{{n\choose 2}}{4^k}\, \l(1 - \frac{1}{2^k}\r)^{n}\r)^m \ge \;\sum_{m=1}^{\infty} %
\sum_{k=1}^{\lceil \alpha^{j}\rceil} \l(\frac{1}{2}\, \frac{n^2}{2}\, %
\frac{\exp{\l(-\frac{1}{\sqrt{n}}\r)}}{4^k}\, \l(1-\frac{1}{2^k}\r)^{n}\r)^{m}.
\ee

\no Now, $1 - x\ge \exp{(-x - x^2)}$ when $x \in \l[0,1/2\r]$, and
\ben
t'_{j} & \ge & \sum_{m=1}^{\infty} e^{-\frac{m}{\sqrt{n}}}\; %
\sum_{k=1}^{\lceil \alpha^{j}\rceil} \frac{1}{4^m} %
\l(\frac{n^2}{4^k}\; \exp{\l(-\frac{n}{2^k} - \frac{n}{4^k}\r)}\r)^{m} \cr
& & \cr
& \ge & \sum_{m=1}^{\infty} e^{-\frac{m}{\sqrt{n}}}\; %
\sum_{k=r}^{\lceil \alpha^{j} \rceil} \frac{1}{4^m}  %
\l(\frac{n^2}{4^k}\; \exp{\l(-\frac{n}{2^k} - \frac{n}{4^r}\r)}\r)^{m}, \label{innersum}
\een
where the last summation is starting from $r\equiv r(n) = 3/2 \log_{2}(n)$.
For such a choice of $r$, we have $2^r \ll n \ll 4^r$ (which is used
in the following).

\mm Therefore, we can now use Lemma~\ref{LEMME-MELLIN2}
to deal with the inner sum in Eq.~\eref{innersum}
\be
t'_{j} & \ge & \sum_{m=1}^{\infty} \exp{\l(-\frac{m}{\sqrt{n}} - m \frac{n}{4^r}\r)} %
\;\sum_{k=r}^{\lceil \alpha^{j} \rceil} \frac{1}{4^m}\, %
\l(\frac{n^2}{4^k}\; \exp{\l(-\frac{n}{2^k}\r)}\r)^m \cr
& & \cr
& \ge & \sum_{m=1}^{n^{1/4}} e^{-\frac{m}{\sqrt{n}} \,-\, \frac{m n}{4^r}} %
\,\l(\frac{(2m-1)!}{4^m\,m^{2m+1}\,\ln{2}} \,-\, %
\frac{1}{m^2}\, \sup_{\nu}|V_{\nu}(\log_2{n})|\r) \;+\; o(1).
\ee
By Lebesgue's monotone convergence theorem, this sum converges to
\be
\sum_{m=1}^{+\infty} \l(\frac{(2m-1)!}{4^m\;m^{2m+1}\;\ln{2}}%
\;-\; \frac{1}{m^2}\, \sup_{\nu}|V_{\nu}(\log_2{n})|\r),
\ee
and the numerical value obtained is $\dis \liminf_n t'_{j} \,\ge \, .39856\ldots$

Finally, from the lower bounds values on $s'_{j}$ and $t'_{j}$, we find
$$\liminf_n p'_{j} \;\ge \;\PDEUXSTAR4\ldots$$

\begin{rem} \label{alpha-cq2}
The lower bound $p_2^{\star}$ in Algorithm~2 is already defined
in~(\ref{CC}). Now, since $p_2^{\star} = \PDEUXSTAR$, $\alpha$ meets
the new condition if it belongs to $(1,1.086\ldots)$, and the minimum
value of the constant $\mathcal{C}(p_2^{\star},\alpha)$ is
$\mathcal{C}(p_2^{\star},\widetilde{\alpha}) \simeq 52.516$, with
$\widetilde{\alpha} = 1.0404\ldots$
\end{rem}

\no Note also that Algorithms~1 and 2 can be improved by starting from
$k = k_0$, $k_0 > 1$ (instead of $1$) in the third line of the
algorithms. Though the running time of each algorithm remains
asymptotically the same, starting from $k = k_0 > 1$ reduces the
awake time to $(1 + \epsilon) \log_{\alpha}\log_{2}(n)$ time slots
for Algorithm~1, and $(1.5 + \epsilon) \log_{\alpha}\log_{2}(n)$
time slots for Algorithm~2 (with $\epsilon = 1/2^{k_0-1}$).
Yet, this also makes the running time longer for small values of $n$;
whence the (obvious) fact that the knowledge of any lower bound
on $n$ greatly helps.

\appendix
\section{Appendix} \label{TECHLEM}
The following two technical Lemmas are at the basis of the asymptotic
complexity analysis of Algorithm~1 or Algorithm~2, in the proof of
Theorem~\ref{Theorem-ALGO1} and Theorem~\ref{Theorem-ALGO2},
respectively. They both use Mellin transform
techniques~\cite{Philippe-Mellin1,PHILIPPE-ROBERT,KNUTH}

\begin{lem} \label{LEMME-MELLIN1}
Let $r\equiv r(n)$ such that, when $n\to +\infty$ we have simultaneously
$r \to \infty$ and $n/2^{r} \to 0$. Then, for all positive integer $m$,
\ben
\sum_{k={1}}^{{r}} \l(\frac{n}{2^k}\r)^m %
\exp{\l(-\frac{nm}{2^k}\r)} & = & \frac{m!}{m^{m+1}\, \ln{2}} %
\;+\; \frac{1}{m\, 2^m}\; U_{m}\B(\log_{2}{(n)}\B) \cr
& & \cr
& + & \BO\l(\frac{2^m}{n^m}\r) \;+\; \BO\l(\frac{n^m}{2^{rm}}\r). \label{EQ:LEMME-MELLIN1}
\een
Denote $\dis \chi_{\ell}\equiv 2i\ell \pi/\ln{2}$.
For any positive integer $m$, $U_{m}\B(\log_{2}{(n)}\B)$ is defined as
\beq
U_{m}(z) =\; \frac{-2^m}{m^{m-1}\, \ln{2}}\, %
\sum_{\ell \in {\qZ}\setminus \{0\}} \Gamma(m + \chi_{\ell}) %
\exp{\B(-\chi_{\ell} \ln(m)\B)} \exp{(-2i\ell \pi z)}. \label{EQ:UMZPHI}
\eeq
The Fourier series $U_{m}(z)$ has mean value $0$ and
the amplitude of its coefficients does not exceed $\FLUCTUN$.
\end{lem}

\begin{proof}
The asymptotic approximation of the finite sum in Eq.~\eref{EQ:LEMME-MELLIN1}
is obtained by direct use of the properties of the Mellin
transform~\cite{Philippe-Mellin1,PHILIPPE-ROBERT}.

The sum
$$f(x) =\; \sum_{k={1}}^{{+\infty}} \l(\frac{x}{2^k}\r)^m \exp{\l(-\frac{mx}{2^k}\r)}$$
is to be analyzed as $x\to \infty.$
Its Mellin transform, with fundamental strip $\langle -1,0\rangle$, is
$$f^{*}(s) =\; \frac{\Gamma(s+m)2^{s}}{m^{s+m}(1-2^{s})}\,.$$

\no There is a simple pole at $s = 0$, but also simple imaginary poles
at $s = \chi_{\ell}\equiv 2 i \ell \pi/\ln{2}$\ (for all non zero integers
$\ell$), which are expected to introduce periodic fluctuations.
The singular expansion of $f^{*}(s)$ in $\langle -1/2,2\rangle$ is
$$f^{*}(s) \asymp \;\l[ \frac{\Gamma(m)}{m^{m}\ln{2}}\, \frac{1}{s} \r] %
\;-\; \frac{1}{m^{m}\ln{2}} \sum_{\ell \in {\qZ}\setminus \{0\}} %
\frac{\Gamma(m + \chi_{\ell})\, \exp{\B(-\chi_{\ell}\, \ln(m)\B)}}{s - \chi_{\ell}}\,.$$
Accordingly, by reinserting $\dis \sum_{k={1}}^{r} \l(\frac{x}{2^k}\r)^m %
\exp{\l(-\frac{mx}{2^k}\r)}$ into $f(x)$, one finds the results stated
in Eqs.~\eref{EQ:LEMME-MELLIN1}) and (\ref{EQ:UMZPHI}).
In Eq.~\eref{EQ:LEMME-MELLIN1}, the periodic fluctuations
are occurring under the form of the Fourier series
$U_{m}\B(\log_{2}{(n)}\B)$ with mean value $0$. Besides, for
any positive integer $m$, the maximum of the amplitude of
the Fourier series is taken at $m = 11$ and it is rather small:
\begin{equation} \label{fluc}
\forall m > 0\ \quad |U_{m}(z)|\, \le \sum_{\ell \in {\qZ}\setminus \{0\}} %
\frac{2^m \; |\Gamma(m + \chi_{\ell})|}{m^{m-1}\ln{2}} \;< \FLUCTUN
\end{equation}
The Fourier coefficients also decrease very fast (see e.g.,~\cite{PHILIPPE-ROBERT}
or~\cite[p.~131]{KNUTH}). The error terms $\BO(2^{m}/n^m)$ and $\BO(n^m/2^{r m})$
in Eq.~\eref{EQ:LEMME-MELLIN1} result from the ``truncated'' summation:
$1\le k\le r$. (The Mellin transform itself results in a $\BO\l(n^{-1}\r)$
error term.)
\end{proof}

\bb
\begin{lem} \label{LEMME-MELLIN2}
Again, let $r\equiv r(n)$ such that, when $n \to +\infty$,
$r \to +\infty$ and $n/2^{r} \to 0$. Then, for all positive integer $m$,
\ben
\sum_{k={1}}^{{r}} \frac{1}{4^{m}} \l(\frac{{n^{2}}}{4^{k}}\r)^m %
\exp{\l(-\frac{nm}{2^k}\r)} & = & \frac{(2m-1)!}{4^{m} m^{2m+1}\,\ln{2}} %
\;+\; \frac{1}{m^2}\; V_{m}\B(\log_{2}{(n)}\B) \cr
& & \cr
& + & \BO\l(\frac{2^{m}}{n^m}\r) \;+\; \BO\l(\frac{n^m}{2^{{r}m}}\r).
\label{EQ:LEMME-MELLIN2}
\een
Again denote $\dis \chi_{\ell}\equiv 2i\ell \pi/\ln{2}$.
For any positive integer $m$, the above Fourier series
$$V_{m}(z) =\; - \frac{m}{4^{m} m^{2m}\, \ln{2}}\, %
\sum_{\ell \in {\qZ}\setminus \{0\}} \Gamma(2m + \chi_{\ell})\, %
\exp{\B(-2i\ell \pi \log_{2}(m)\B)}\, \exp{(-2i\ell \pi z)}$$
has mean value $0$ and the amplitude of the coefficients
of $V_m(z)$ cannot exceed $\FLUCTDEUX$.
\end{lem}

\begin{proof}
The asymptotic approximation of the finite sum in Eq.~\eref{EQ:LEMME-MELLIN2}
is provided along the same lines as in Lemma~\ref{LEMME-MELLIN1}. Now,
$$g(x) =\; \sum_{k={1}}^{+\infty} \frac{1}{4^{m}} %
\l(\frac{x}{2^k}\r)^{2m} \exp{\l(-\frac{mx}{2^k}\r)}$$
is to be analyzed when $x\to \infty$. Similarly, its Mellin transform is\
$\dis g^{*}(s) = \frac{\Gamma(s+2m)2^{s}}{m^{s+2m}(1-2^{s})}\,$,
with fundamental strip $\langle -1,0\rangle$. Again there is a simple pole
at $s = 0$\ and simple imaginary poles at $s = \chi_{\ell}\equiv 2i\ell \pi/\ln{2}$\
(for all non zero integers $\ell$), which are also expected
to introduce periodic fluctuations.

As in Lemma~\ref{LEMME-MELLIN1} the singular expansion of
$g^{*}(s)$ provides an asymptotic approximation of the finite
sum in Eq.~(\ref{EQ:LEMME-MELLIN2}). The periodic fluctuations occur
under the form of the Fourier series $V_{m}\B(\log_{2}{(n)}\B)$.
As $U_{m}(z)$, $V_{m}(z)$ has mean value $0$ and its coefficients
have a very tiny amplitude. Their minimum is taken at $m = 2$, and
$$\forall m > 0\ \quad |V_{m}(z)| \,\le \sum_{\ell \in {\qZ}\setminus \{0\}} %
\frac{|\Gamma(2m + \chi_{\ell})|}{4^m\, m^{2m-1}\,\ln{2}} \;<\, \FLUCTDEUX.$$
The Fourier coefficients also decrease very fast (see e.g.,
\cite{PHILIPPE-ROBERT} or~\cite[p.~131]{KNUTH}). Last, the ``truncated''
summation ($1\le k\le r$) results again in error terms $\BO(2^{m}/n^m)$
and $\BO(n^m/2^{r m})$ in Eq.~\eref{EQ:LEMME-MELLIN2}.
\end{proof}

\end{document}